\documentclass[fleqn,10pt]{wlscirep}

\usepackage{graphicx,epsfig}

\usepackage{times}
\usepackage{graphics,dcolumn,bm,fleqn,float}
\usepackage{amssymb,amsmath,multirow,rotate,color}
\usepackage{comment} 
\usepackage{cite} 
\usepackage{stackengine}

\DeclareMathAlphabet{\altmathcal}{OMS}{cmsy}{m}{n}

\graphicspath{{./figures/}}

\title{Online division of labour: emergent structures in Open Source Software}

\author[1]{Mar\'ia J.~Palazzi}
\author[1,2]{Jordi Cabot}
\author[1]{Javier Luis C\'anovas Izquierdo}
\author[1]{Albert Sol\'e-Ribalta}
\author[1]{Javier Borge-Holthoefer}
\affil[1]{Internet Interdisciplinary Institute (IN3), Universitat Oberta de Catalunya, Barcelona, Catalonia, Spain}
\affil[2]{ICREA, Barcelona, Catalonia, Spain}

\begin{abstract}
The development Open Source Software fundamentally depends on the participation and commitment of volunteer developers to progress on a particular task. Several works have presented strategies to increase the on-boarding and engagement of new contributors, but little is known on how these diverse groups of developers self-organise to work together. 
To understand this, one must consider that, on one hand, platforms like GitHub provide a virtually unlimited development framework: any number of actors can potentially join to contribute in a decentralised, distributed, remote, and asynchronous manner. On the other, however, it seems reasonable that some sort of hierarchy and division of labour must be in place to meet human biological and cognitive limits, and also to achieve some level of efficiency.
These latter features (hierarchy and division of labour) should translate into recognisable structural arrangements when projects are represented as developer-file bipartite networks. 
Thus, in this paper we analyse a set of popular open source projects from GitHub, placing the accent on three key properties: nestedness, modularity and in-block nestedness --which typify the emergence of heterogeneities among contributors, the emergence of subgroups of developers working on specific subgroups of files, and a mixture of the two previous, respectively.
These analyses show that indeed projects evolve into internally organised blocks. Furthermore, the distribution of sizes of such blocks is bounded, connecting our results to the celebrated Dunbar number both in off- and on-line environments.
Our analyses create a link between bio-cognitive constraints, group formation and online working environments, opening up a rich scenario for future research on (online) work team assembly (e.g. size, composition, and formation). From a complex network perspective, our results pave the way for the study of time-resolved datasets, and the design of a suitable model that can mimic the growth and evolution of OSS projects.

The development Open Source Software fundamentally depends on the participation and commitment of volunteer developers to progress. Several works have presented strategies to increase the on-boarding and engagement of new contributors, but little is known on how these diverse groups of developers self-organise to work together. To understand this, one must consider that, on one hand, platforms like GitHub provide a virtually unlimited development framework: any number of actors can potentially join to contribute in a decentralised, distributed, remote, and asynchronous manner. On the other, however, it seems reasonable that some sort of hierarchy and division of labour must be in place to meet human biological and cognitive limits, and also to achieve some level of efficiency. These latter features (hierarchy and division of labour) should translate into recognisable structural arrangements when projects are represented as developer-file bipartite networks. In this paper we analyse a set of popular open source projects from GitHub, placing the accent on three key properties: nestedness, modularity and in-block nestedness -which typify the emergence of heterogeneities among contributors, the emergence of subgroups of developers working on specific subgroups of files, and a mixture of the two previous, respectively. These analyses show that indeed projects evolve into internally organised blocks. Furthermore, the distribution of sizes of such blocks is bounded, connecting our results to the celebrated Dunbar number both in off- and on-line environments. Our analyses create a link between bio-cognitive constraints, group formation and online working environments, opening up a rich scenario for future research on (online) work team assembly. 

\end{abstract}

\begin{document}

\maketitle

\section*{Introduction}

Open Source Software (OSS) is a key actor in the current software market, and a major factor in the consistent growth of the software economy. The promise of OSS is better quality, higher reliability, more flexibility, lower cost, and an end to predatory vendor lock-in, according to the Open Source initiative\cite{opensource}. These goals are achieved thanks to the active participation of the community\cite{SchuwerGH15}: indeed, OSS projects depend on contributors to progress\cite{Dabbish2012,Padhye2014}. 

The emergence of GitHub and other platforms as prominent public repositories, together with the availability of APIs to access comprehensive datasets on most projects' history, has opened up the opportunities for more systematic and inclusive analyses of how OSS communities operate. 
In the last years, research on OSS has left behind a rich trace of facts. For example, we now know that the majority of code contributions are highly skewed towards a small subset of projects\cite{Lima2014,Dabbish2013}, with many projects quickly losing community interest and being abandoned at very early stages\cite{Fitz-Gerald12b}. Moreover, most projects have a low {\it truck factor}, meaning that a small group of developers is responsible for a large set of code contributions\cite{CosentinoIC15,Yamashita2015,Avelino2016}. This pushes projects to depend more and more on their ability to attract and retain occasional contributors (also known as ``drive-by'' commits\cite{Pham2013}) that can complement the few core developers and help them to move the project forward. Along these lines, several works have focused on strategies to increase the on-boarding and engagement of such contributors (e.g., by using simple contribution processes\cite{Yamashita2016}, extensive documentation\cite{Hata2015}, gamification techniques\cite{BertholdoG16} or {\it ad hoc} on-boarding portals\cite{SteinmacherCTG16}, among others\cite{SteinmacherSGR15}). Other social, economic, and geographical factors affecting the development of OSS have been scrutinised as well, see Cosentino {\it et al.}\cite{cosentino2017systematic} for a thorough review.

Parallel to these macroscopic observations and statistical analyses, social scientists and complex network researchers have focused, in relatively much fewer papers, on analysing how a diverse group of (distributed) contributors work together, i.e. the structural features of projects. Most often, these works pivot on the interactions between developers, building explicit or implicit collaborative networks, e.g. email exchanges\cite{valverde2007self,bird2008latent} and unipartite projections from the contributor-file bipartite network\cite{hong2011understanding}, respectively. These developer social networks have been analysed to better understand the hierarchies that emerge among contributors, as well as to identify topical clusters, i.e. cohesive subgroups that manifest strongly in technical discussions. However, the behaviour of OSS communities cannot be fully understood only accounting for the relations between project contributors, since their interactions are mostly mediated through the edition of project files (no direct communication is present between group members). To overcome this limitation, we focus on studying the structural organisation of OSS projects as contributor-file bipartite graphs. Far beyond technical and methodological adaptations, the consideration of these two elements composing the OSS system allows retaining valuable information (as opposed to collapsing it on a unipartite network) and, above all, recognising both classes as co-evolutionary units that place mutual constraints on each other.

Our interest on the structural features of OSS projects departs from some obvious, but worth highlighting, observations. First, public collaborative repositories place no limits, in principle, to the number of developers (and files) that a project should host. In this sense, platforms like GitHub resemble online social networks (e.g. Twitter or Facebook), in which the number of allowed connections is virtually unbounded. However, we know that other factors --biological, cognitive-- set well-defined limits to the amount of active social connections an individual can have\cite{dunbar1992neocortex}, also online\cite{gonccalves2011modeling}. But, do these limits apply in collaborative networks, where contributors work remotely and asynchronously? Does a division of labour arise, even when interaction among developers is mostly indirect (that is, via the files that they edit in common)? And, even if specialised subgroups emerge (as some evidence already suggests, at least in developer social networks\cite{hong2011understanding}), do these exhibit some sort of internal organisation?

To answer these questions, we will look at three structural arrangements which have been identified as signatures of self-organisation in both natural and artificial systems: nestedness\cite{patterson1986patterson,atmar1993measureorder} (i.e. do projects evolve in a way such that the emergence of generalists and specialists is favoured?); modularity\cite{newman2004finding,barber2007modularity,fortunato2010community} (i.e. do OSS projects split in identifiable compartments, thus avoiding Brook's law\cite{Brooks1995} despite the addition of contributors? Are these compartments bounded?); and in-block nestedness\cite{lewinsohn2006structure,flores2013multi,sole2018revealing} (i.e. if bio-cognitive limits and division of labour are in place, do the resulting specialised modules self-organise internally?).



\section*{Results}
The projects that we analyse in the following were selected according to their popularity (quantified as the number of stars these projects had received on GitHub, at the time of collection in 2016). This criterium mainly responds to two arguments: maturity and success. That is, here we purposefully pay attention to projects which have reached a reasonable degree of evolution, regardless of the absence (or presence) of any given structural organisation at the initial stages. 

After pre-processing, formatting and discarding some of the top 100 public OSS projects hosted on GitHub, we ended up retaining 65 of them, see Materials and Methods for details. As can be seen in Table~\ref{table1}, we have a sufficiently broad distribution of project sizes and age. Note also that popularity (number of stars) is not necessarily related to their size or age. Each of these projects have been represented as a bipartite unweighted graph, where inter-class links (between contributors $c$ and files $f$) are allowed, but intra-class links are forbidden. This bipartite network is thus encoded as an $N \times M$ rectangular binary matrix $\mathbf{A}$, where entries $a_{cf} = 1$ if contributor $c$ edited the file $f$ at least once. The size of a project is $S=N+M$; the smallest project considered here is \textit{resume.github.com}, with $S=82$, and the largest one is \textit{foundation-sites}, with $S=13,382$.

\begin{table}[ht]
\caption{Statistics of our dataset.}
\begin{center}
\begin{tabular}{c|c|c|c|c|c}
 & \# contributors & \# files & \# commits & \# stars & Project age \\\hline
Largest project & 1,061 & 12,321& 75,757 & 27,500& 4 years 11 months \\
Smallest project & 55 & 27 & 444 & 36,900 & 5 years 6 months \\
Average  & 422 & 3,247 & 33,936 & 46,334 & 4 years 9 months \\\hline 
Most popular project & 516 & 2,833 & 34,666 & 293,000 & 2 years 10 months \\
Least popular project & 117 & 103 & 4,057 & 21,700 & 5 years 5 months \\\hline
Oldest project & 1,434 & 10,413 & 174,452 & 35,000 & 11 years 3 months \\
Youngest project & 43 & 51 & 210 & 31,600 & 0 years 3 months \\
\end{tabular}
\end{center}
\label{table1}
\end{table}%

\begin{figure}[ht]
\centering
\includegraphics[width=0.5\textwidth]{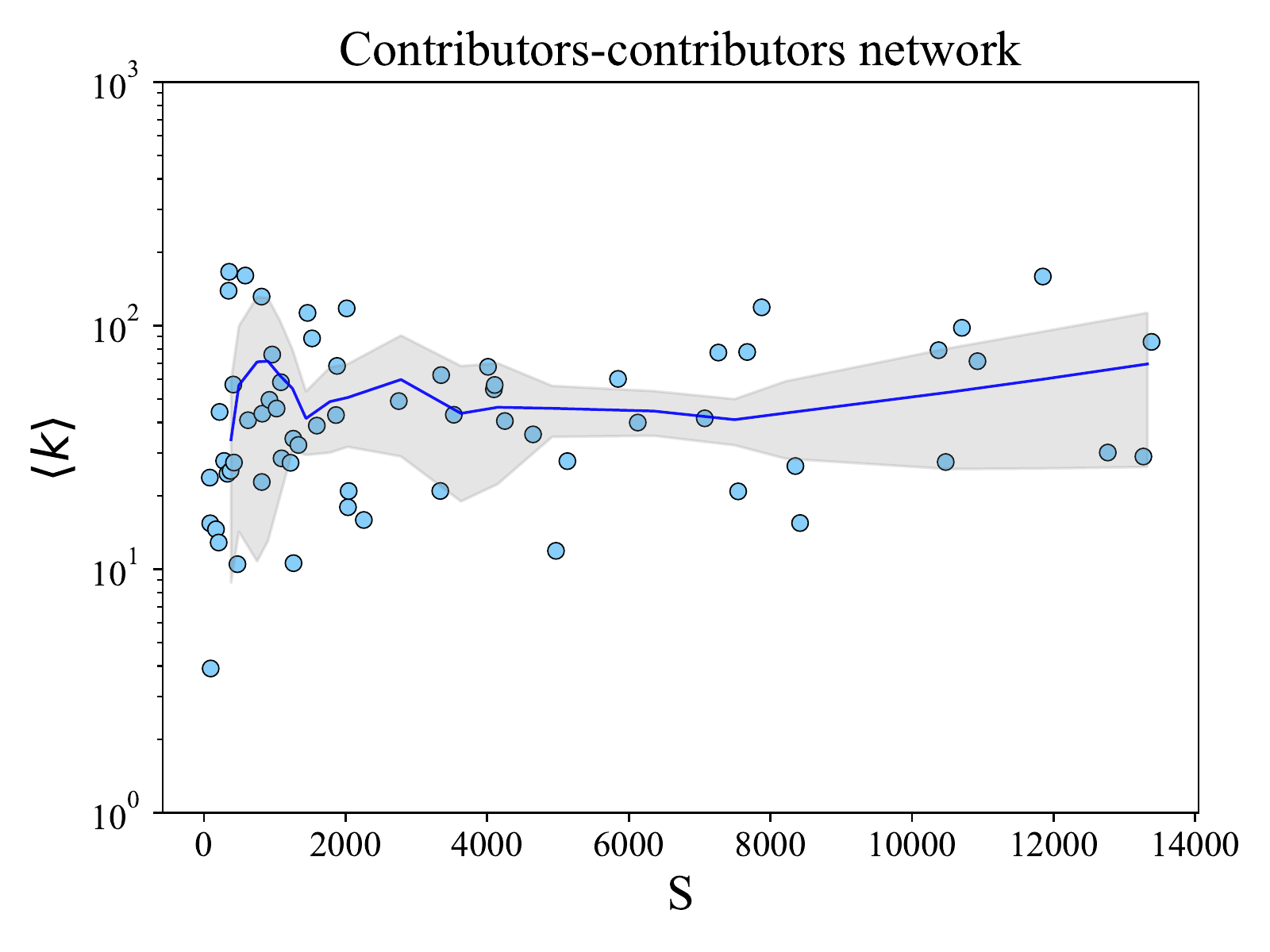}
\caption{Scatter plot of the developers implicit average degree $\langle k \rangle$ (blue line) against the size $S=N+M$ of a project (where $N$ is the number of contributors, and $M$ is the number of files. The shadowed grey area represents one standard deviation above and below the average, while circles represent each individual project. The plot is presented in semi-log.}
\label{fig_k_avg}
\end{figure}

\subsection*{Preliminary observations}
Before we focus on the structural arrangements of interest (nestedness, modularity, in-block nestedness), we explore whether a potentially unbounded interaction capability is mirrored in actual OSS projects across 4 orders of magnitude in size. To do so, we work on the projected contributor-contributor network, to measure the developer's implicit average degree $\langle k \rangle$, i.e. the average amount of contributors with whom an individual shares at least one file. Figure \ref{fig_k_avg} shows a scatter plot of $\langle k \rangle$ against $s$ (note the semi-log scaling). Besides the initial fluctuating pattern, it is clear that $\langle k \rangle$ presents an almost flat trajectory which indicates that, on average, a contributor indirectly interacts with $\sim70$ peers, regardless of the size of the project.

\begin{figure}[ht]
\centering
\hspace{.1cm} \includegraphics[width=0.82\textwidth]{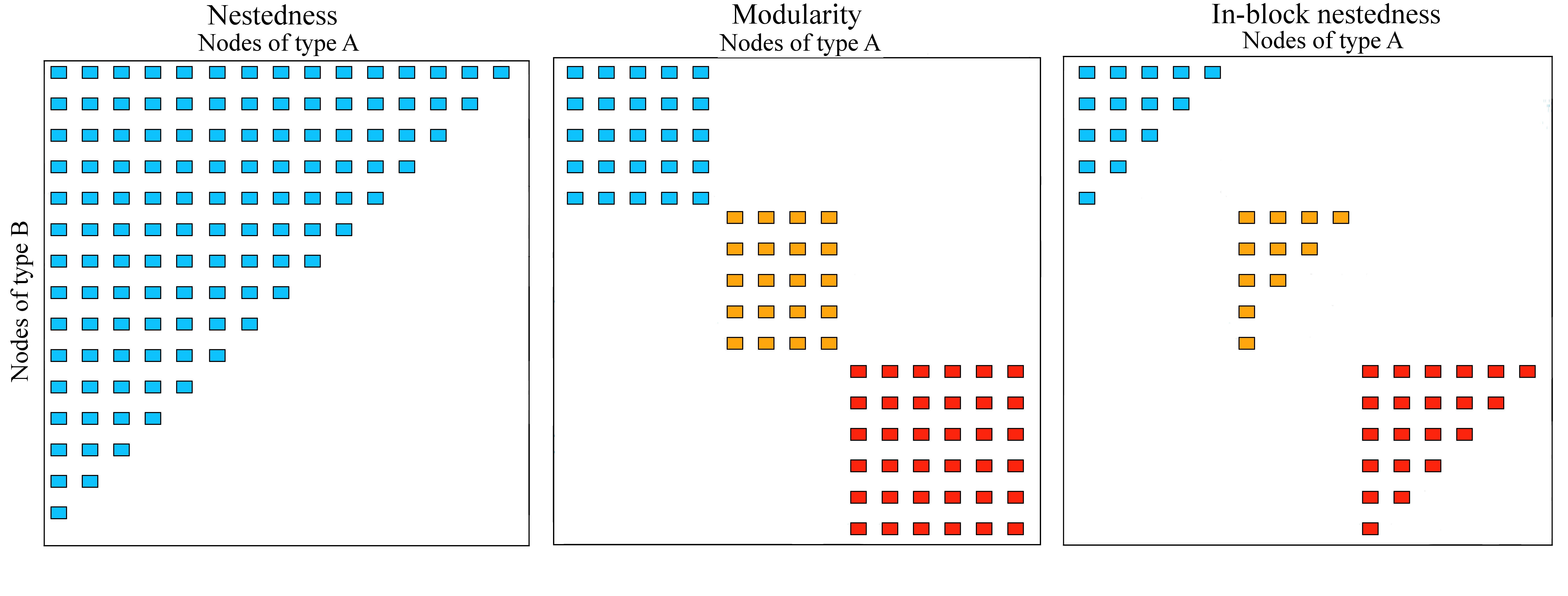}\\ \vspace{-.5cm}
 \includegraphics[width=0.82\textwidth]{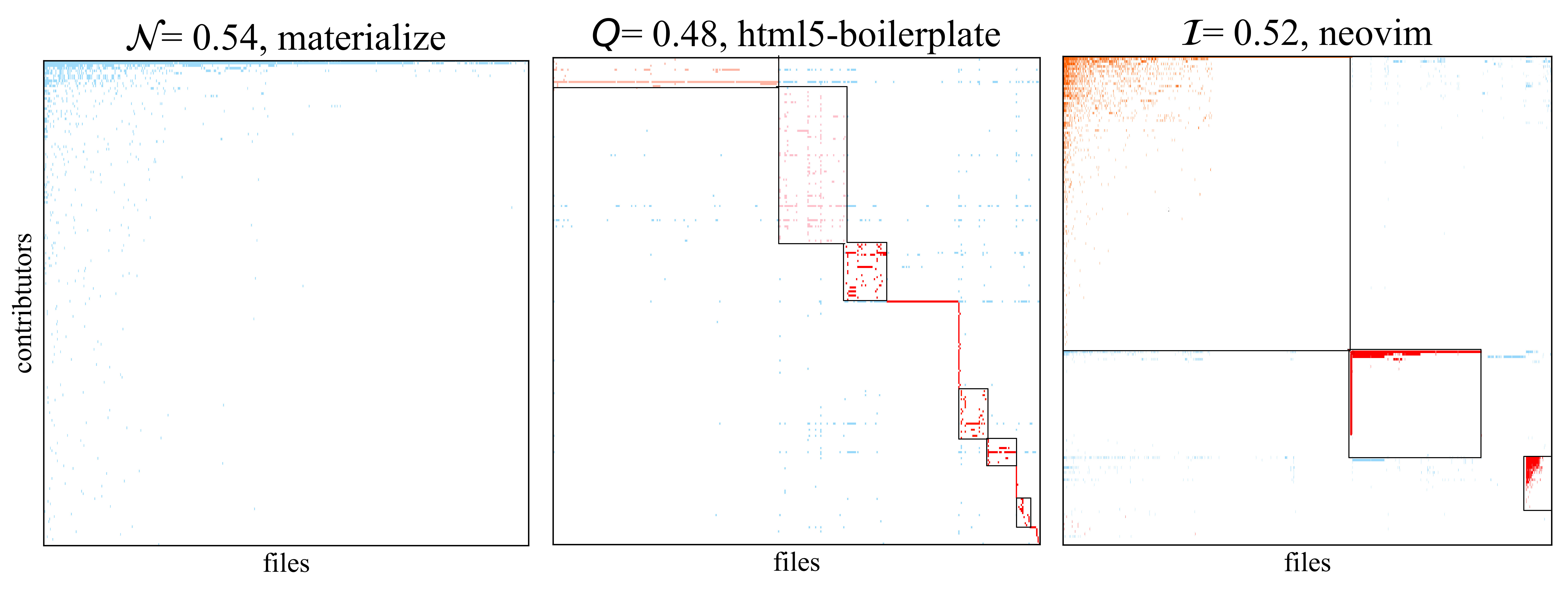}\hspace{-.2cm}
        \caption{Top row: left: Nestedness $\altmathcal{N}$, middle: Modularity $Q$, bottom: In-block nestedness $\altmathcal{I}$. Bottom row:  Interaction matrices for three projects with high values for each one the structural patterns of interest.}
        \label{fig_structures_matrices}
\end{figure}

The flat pattern for the developers implicit average degree in Figure~\ref{fig_k_avg} is interesting in two aspects. First, it points to an inherent limitation to the number of connections (even indirect ones) that a contributor to a project can sustain. Notably, such limitation is below (but not far) from the celebrated Dunbar number (somewhere between 100 and 300), which is echoed as well in digital environments\cite{gonccalves2011modeling}. Second, the result is consistent with the existence of some sort of mesoscale organisation in the projects. In Bird {\it et al.}\cite{bird2008latent}, the authors find that developers in the same community have more files in common than pairs of developers from different communities. Reversing the argument, one may say that relatively small contributor neighbourhoods are indicative, though not a guarantee, of the presence of well-defined subgroups in OSS projects.

\subsection*{Mesoscale patterns}
From the previous encouraging result, we move on to the analysis of the projects at a larger scale. The specificities of the methods to calculate nestedness $\altmathcal{N}$, and to optimise modularity $Q$ and in-block nestedness $\altmathcal{I}$ are detailed in the Materials and Methods section. For the sake of illustration, Figure~\ref{fig_structures_matrices} (top row) shows idealized examples of nestedness (left), modular (middle) and in-block nested (right) arrangements. The bottom row of the figure presents actual adjacency matrices of three projects with high values of each one of the structural measures. In them, rows and columns have been rearranged to highlight the different properties.

\begin{figure}[!ht]
\centering
\def\stackalignment{l}
\includegraphics[width=.6\textwidth]{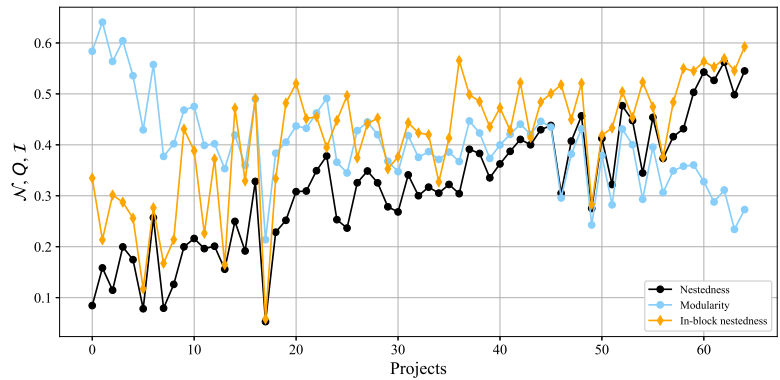}
\caption{$\altmathcal{N}$, $Q$, and $\altmathcal{I}$ obtained values, for each project of our dataset. The projects were sorted to maximise the difference between $\altmathcal{N}$ and $Q$.}
\label{fig1-1}
\end{figure}

We start out with a general overview of the results for the three measures of interest. Figure~\ref{fig1-1} plots the obtained values for $\altmathcal{N}$, $Q$, and $\altmathcal{I}$ over all the projects considered in this work. To ease visualisation, and considering that nestedness and modularity are antagonistic organisations\cite{palazzi2018antagonistic}, projects are sorted to maximise the difference between $\altmathcal{N}$ and $Q$. In general, nestedness is the lowest of the three values at stake, and in-block nestedness is, more often than not, the highest. It can be safely said, thus, that a tendency to self-organise as a block structure is present: 90\% of the projects exhibit either $Q$ or $\altmathcal{I}$ above 0.4, and values beyond 0.5 are not rare. This evidence is compatible with previous results regarding the division of labour: indeed, be them modular or in-block nested, most projects can be split into communities of developers and files, forming subgroups around product-related activities\cite{bird2008latent}

Just like there is virtually no technical limit to the overall size of a project, there is not either an explicit bound to the size that a sub-group should have. And yet, previous theory and evidence suggests that larger communities come at an efficiency cost: the dynamics of a group change fundamentally when they exceed the Dunbar number, which is estimated around 150. While most often the number refers to personal acquaintances, it has been (and still is) applied in the industrial sphere\cite{dunbar2010many}. Applied to the OSS environment, exceedingly small working sub-groups might hamper a project's advance; while too many contributors may not allow the group to converge towards a solution\cite{derex2016partial,derex2018divide}. We explore whether, indeed, size limitations arise in developers sub-groups, as they emerge from either $Q$ or $\altmathcal{I}$ optimisation procedures.  Although partitions are hybrid, i.e. a community has both developers and files, in the following, we will report the community sizes in terms of developers.

\begin{figure}[!ht]
\centering
\def\stackalignment{l}
\topinset{\bfseries(a)}{\includegraphics[width=.4\textwidth]{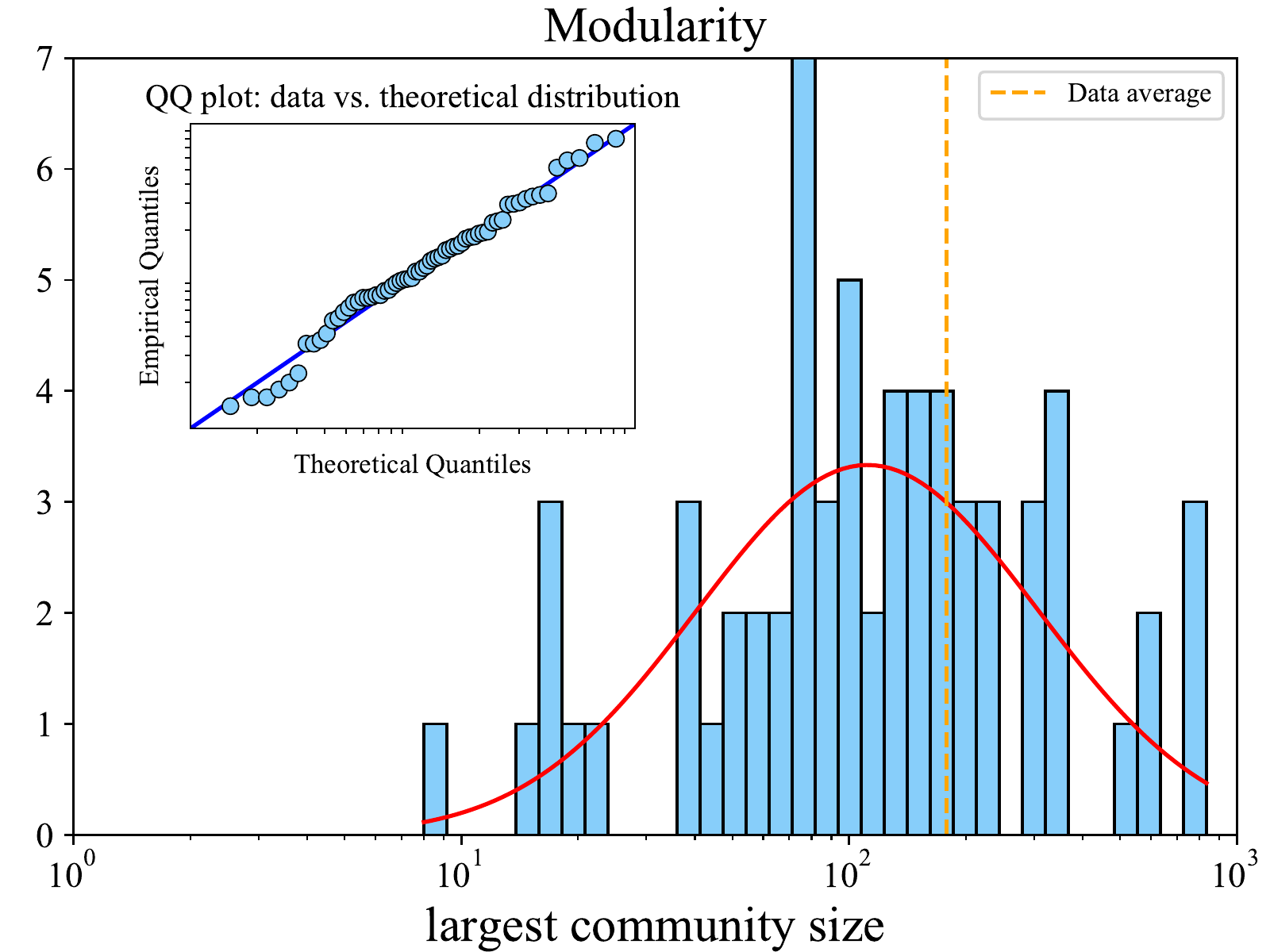}}{-0.1in}{0.05in}
\topinset{\bfseries(b)}{\includegraphics[width=.4\textwidth]{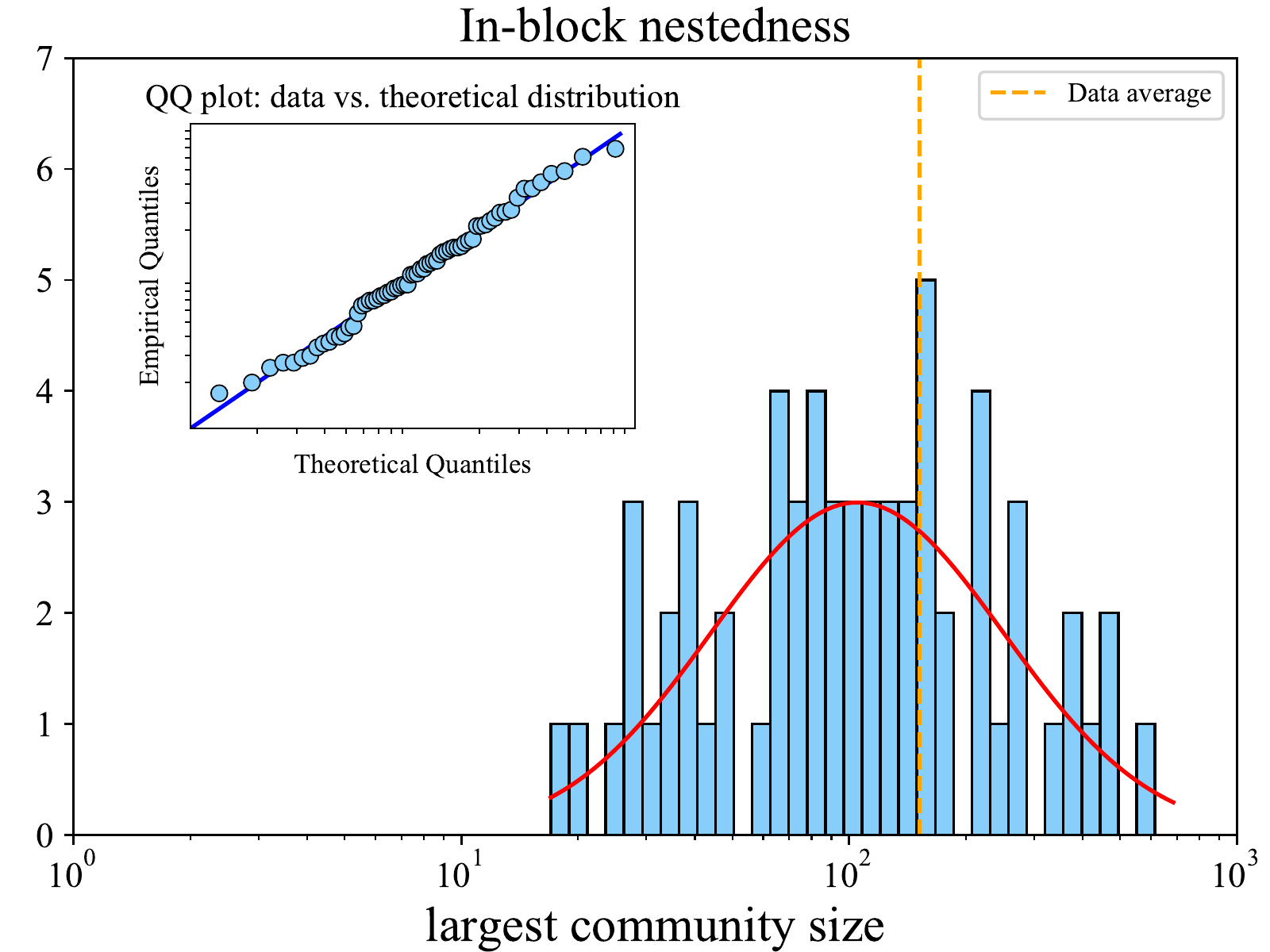}}{-0.1in}{0.05in}
\caption{Distribution of the largest community size for each project obtained after optimization of modularity (panel a) and in-block nestedness (panel b). In both panels, the solid red line corresponds to the log-normal fit performed to each distribution, which are centred around 100. The dashed orange line indicates the average values of our dataset, and inset panels show the Q-Q plots of the empirical versus theoretical quantiles from the log-normal distribution fit.}
\label{fig_scatter_hist}
\end{figure}

Figure~\ref{fig_scatter_hist} provides a global overview of the 65 projects studied here, with the distribution of their largest subgroup sizes as they are identified via $Q$ (panel a) or $\altmathcal{I}$ (panel b). In both cases the average (dashed orange vertical line) is below 200, and the histogram is evenly distributed around 100: most communities belong in the range from 80 to 200. For the sake of comparison, the solid red line represents a log-normal fit (notice the logarithmic scale in the $x$-axis), and the insets in both panels show the Q-Q plots, to compare both theoretical and empirical distributions revealing that, indeed, the fit is accurate.

\begin{figure}[!ht]
\centering
\def\stackalignment{l}
\topinset{\bfseries(a)}{\includegraphics[width=.4\textwidth]{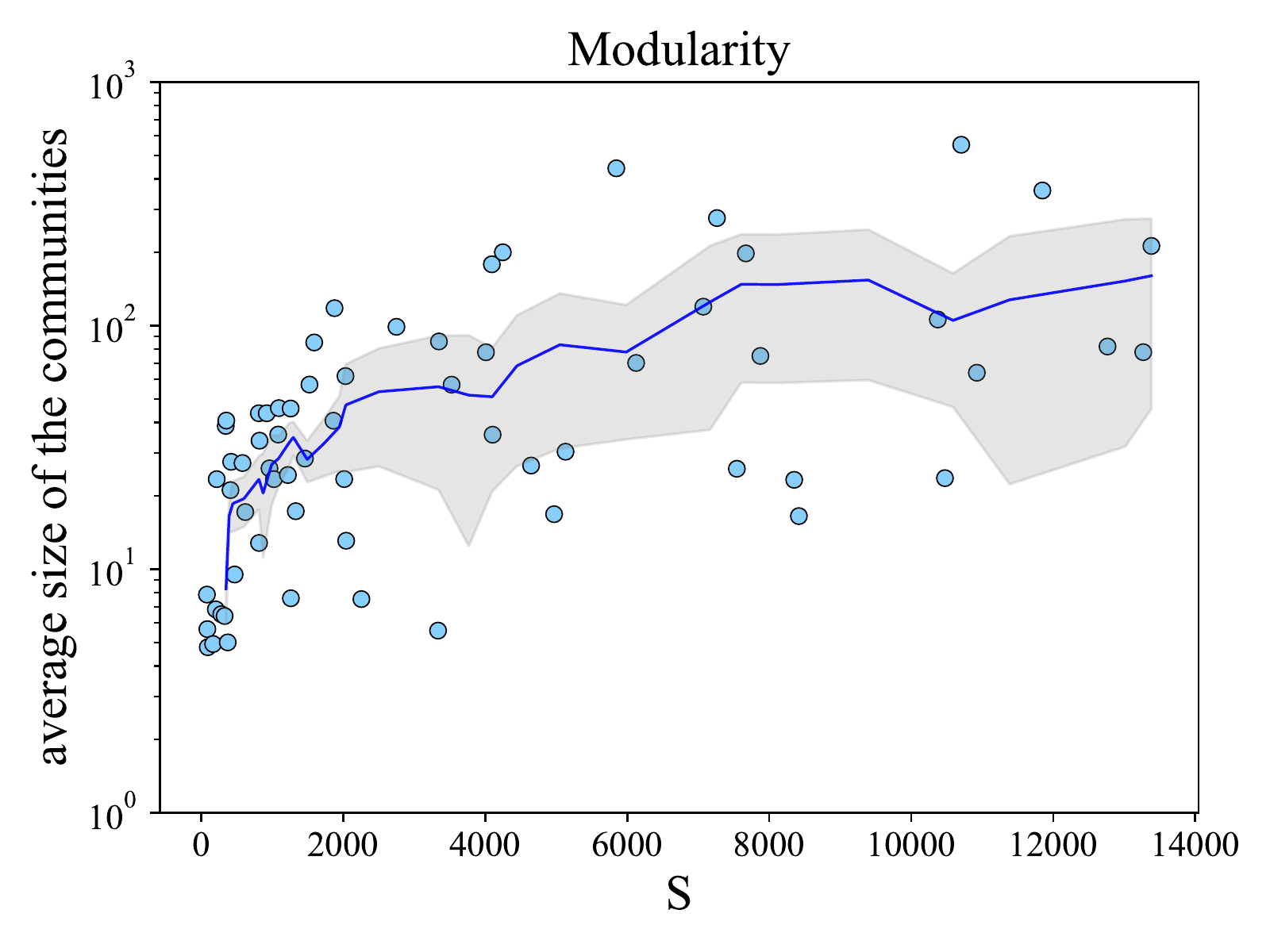}}{-0.1in}{0.05in}
\topinset{\bfseries(b)}{ \includegraphics[width=.4\textwidth]{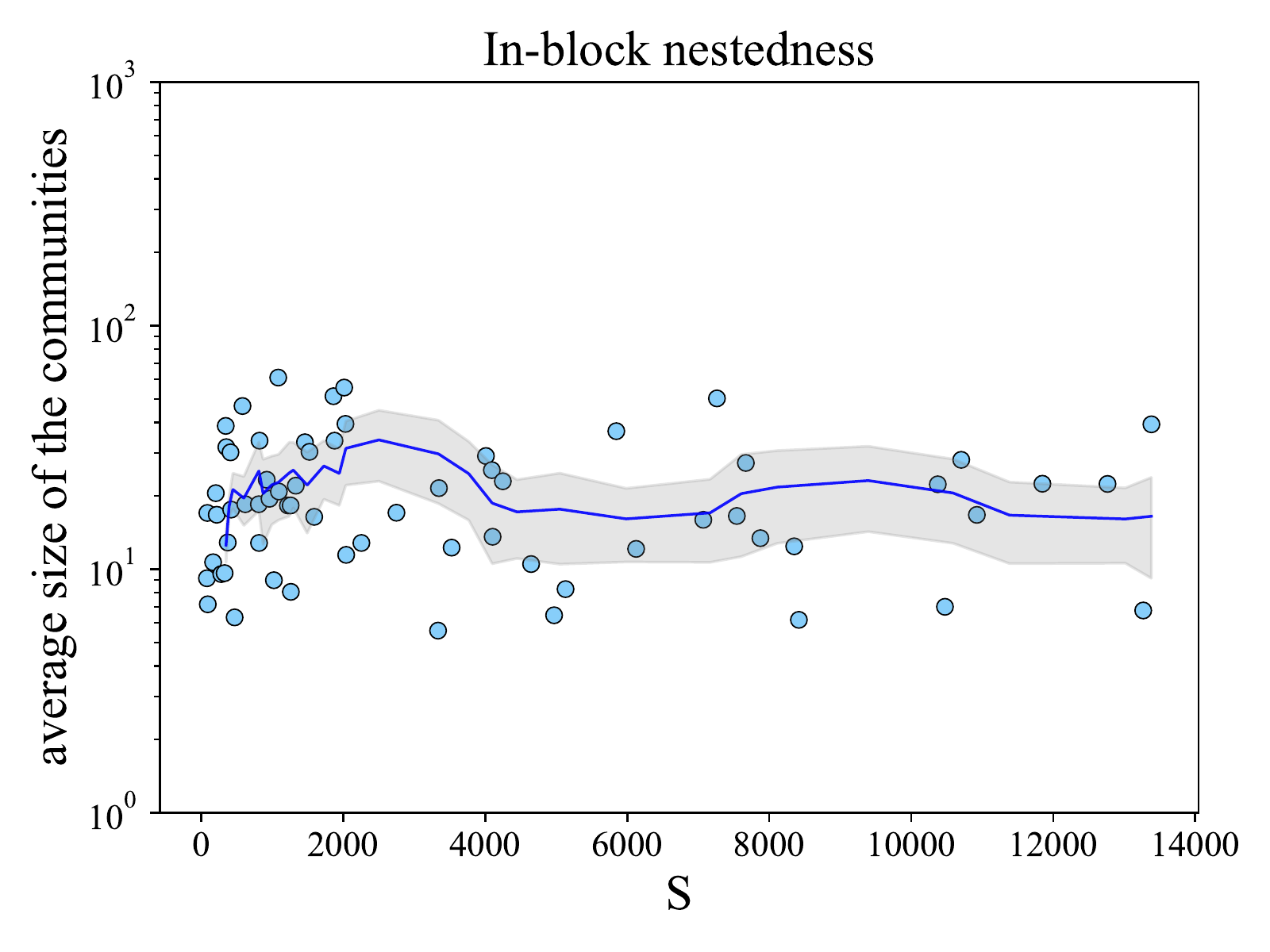}}{-0.1in}{0.05in}
\topinset{\bfseries(c)}{\includegraphics[width=.4\textwidth]{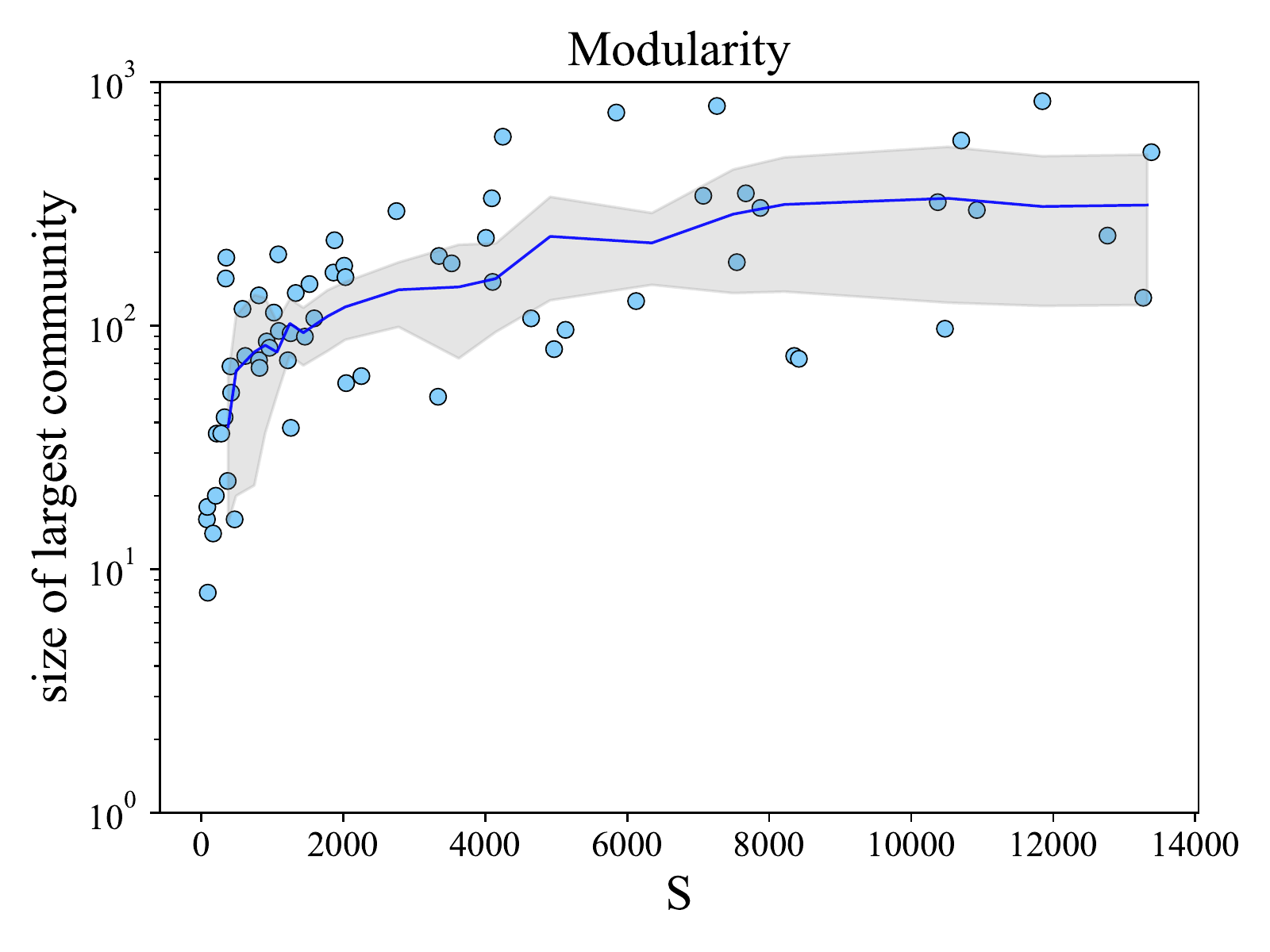}}{-0.1in}{0.05in}
\topinset{\bfseries(d)}{ \includegraphics[width=.4\textwidth]{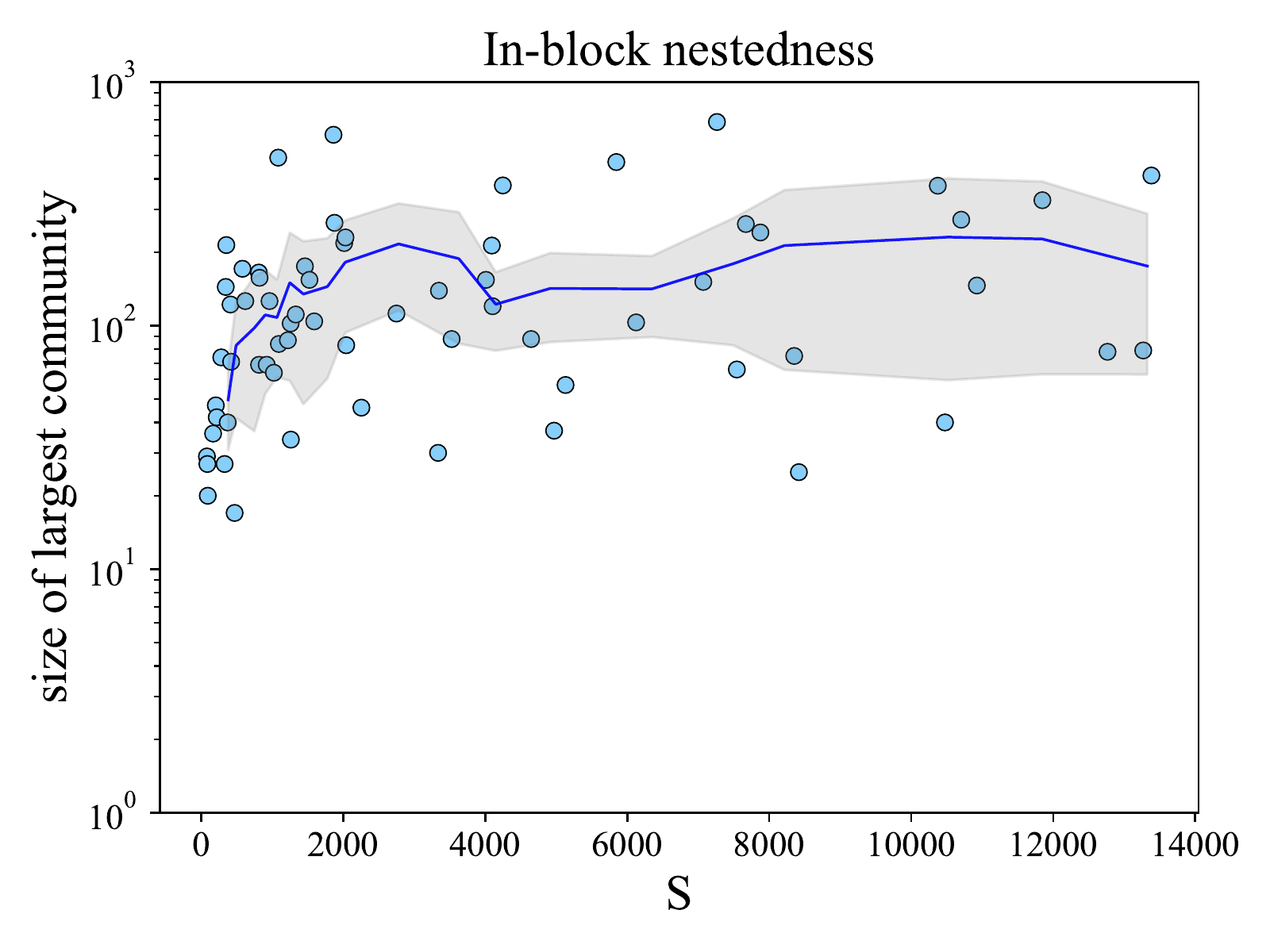}}{-0.1in}{0.05in}
\caption{The evolution of the average community size as a function of $s$ presents differences for $Q$- and $\altmathcal{I}$-optimised partitions (panels a and c, respectively). Regarding the size, average $Q$-communities are in general larger than $\altmathcal{I}$-communities. Furthermore, the scaling behaviour is also different: an average community size for $Q$-optimised partitions moderately grows with $s$, while it remains fairly constant for $\altmathcal{I}$ beyond $s > 2000$. Turning from average to maximum community size, $Q$- and $\altmathcal{I}$-optimised partitions (panels b and d, respectively) present very similar bounds, from 30 to 300 contributors. Again, the largest $Q$-community slightly tends to grow with $s$, while this size stabilises around 100 for the case of $\altmathcal{I}$. Note semi-log scaling.}
\label{fig_maxComm_size}
\end{figure}

Although Figure~\ref{fig_scatter_hist} evidences, on average, a well-defined maximum community size, we must ensure that the size of the largest communities detected for each project is independent of the size of the project, in order to validate such organisational limit. To do so, we go down to the project level. Figure~\ref{fig_maxComm_size} reports average (panels a and c) and maximum (panels b and d) subgroup sizes for both community identification strategies, as a function of the project size $s$. In general, results point at the existence of bounds to group size, which resemble the limits described by Dunbar's number: even the largest projects reflect that the maximum size of a community in them is between 100 and 200 (in the case of $\altmathcal{I}$-communities, panel d). This result is robust and stable beyond $s > 2000$. On the other hand, largest community size is slightly above 200 in the case of $Q$-communities (panel b). These results are ever more striking, since such trend towards the compartmentalisation of the workload is not only decentralised, in the sense that it does not emerge from a predefined plan, but also implicit, because the interaction between developers is most often indirect.

\subsection*{Co-existing architectures and project maturity}
As it has been suggested\cite{palazzi2018antagonistic}, empirical evidence indicates that more than one structural pattern may concur within a network, each evincing different properties of the system. We take the same stance here: a network is not regarded, for example, as {\it completely} modular or {\it completely} nested; rather, it may combine structural features that reflect the evolutionary history of the system, or the fact that the system evolves under different dynamical pressures that favour competing arrangements. 

\begin{figure}[!ht]
\centering
\def\stackalignment{l}
\includegraphics[width=.5\textwidth]{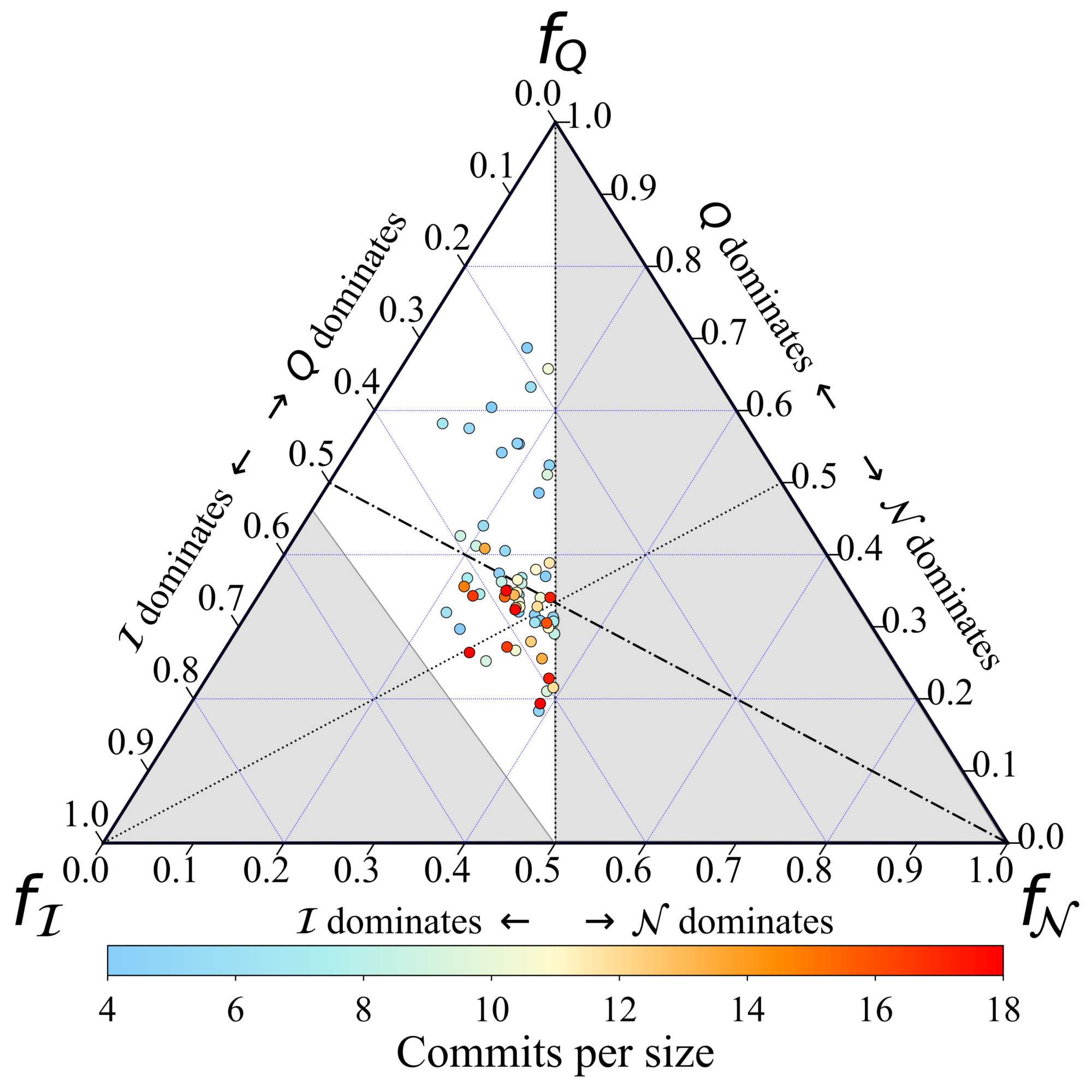}
\caption{Distribution of the three architectural patterns for each the projects across a ternary plot. The colorbar indicates the number of commits received by each project, normalised by the size of it.}
\label{fig_ternary}
\end{figure}

A convenient way to grasp this mixture is a ternary plot (or simplex), see Figure~\ref{fig_ternary}. In the ternary plot, each project is located with three coordinates $f_{\altmathcal{N}}$, $f_{Q}$ and $f_{\altmathcal{I}}$, which are simply calculated from the original scores, e.g. $f_{\altmathcal{N}} = \altmathcal{N} / (\altmathcal{N} + Q + \altmathcal{I})$. The simplex can be partitioned according to ``dominance regions'', bounded by the three angle bisectors. These regions intuitively tell us which of the three patterns is more prominent for any given project. Note that certain areas of the simplex (in grey in Figure~\ref{fig_ternary}) are necessarily empty, see Materials and Methods, and Palazzi {\it et al.}\cite{palazzi2018antagonistic} for further details. Figure~\ref{fig_ternary} reveals that most projects lie in the nested regions, while the predominantly modular region is relatively empty. 

Together with their dominant architecture, points in Figure \ref{fig_ternary} are colour-coded according to the total number of commits that each project has received. We take this number as a proxy to the level of development or maturity of the project (note that a project's age may be misleading due to periods of inactivity). The distribution of colour on the simplex suggests that more mature projects tend to exhibit nested or in-block nested structures, whereas predominantly modular projects appear to be relatively immature (with exceptions, admittedly). Such result is resonant to the fact that topical conversations in online social networks (``information ecosystems") evolve through different stages --modular when the discussion is still brewing in a scattered way; nested when the discussion becomes mainstream to the group of interest\cite{borge2017emergence}. More relevant to OSS development, Figure~\ref{fig_ternary} reconciles the idea of workload compartmentalisation (subcommunities forming around product-related activities)\cite{bird2008latent}, and the emergence of hierarchies\cite{CosentinoIC15} or a rich club\cite{valverde2007self} of developers, at least in well-developed projects. This partial picture is however complemented by the fact that hierarchies emerge as well on the code class: the presence of generalists and specialists applies to both developers and files in a nested or in-block nested scenario.

\section*{Discussion}
In summary, our analyses have unveiled that OSS projects evolve into a relatively narrow set of structural arrangements. At the mesoscale, we observe that projects tend to form blocks, a fact that can be related to the need of contributors to distribute coding efforts, allowing a project to develop steadily and in a balanced way. Focusing on the file class, the emergence of blocks is interesting as well, since a modular architecture (understood now as a software design principle) is a desired feature in any complex software project. 
Furthermore, those blocks or subgroups have a relatively stable size no matter how large a project is. Remarkably, such size fluctuates around the Dunbar number.

Previous research reported that OSS projects are largely heterogeneous, in the sense that developers self-organise into hierarchical structures. Such statement may seem to clash with a modular arrangement, to the extent that modularity $Q$ does not make any assumption regarding the internal organisation of the subgroups. Our findings, however, point at the fact that more mature projects tend to present a nested organisation inside modules. Thus, the presence of workload compartmentalisation is compatible with the emergence of hierarchies, with generalists and specialists throughout a project. Paradoxically, a more evolved and structured architecture does not imply better overall performance here: the nested arrangement inside blocks can hamper a project's progress, since the occasional and least committed contributors --those acting upon a small part of the code-- tend to edit precisely the most generalist files, neglecting the least developed ones.

These findings open up a rich scenario, with many questions lying ahead. On the OSS environment side, our results contribute to an understanding of how successful projects self-organise towards a modular architecture: large and complex tasks, involving hundreds (and even thousands) of files appear to be broken down, presumably for the sake of efficiency and task specialisation (division of labour). Within this compartmentalisation, mature projects exhibit even further organisation, arranging the internal structure of subgroups in a nested way --something that is not grasped by modularity optimisation only. More broadly, our results demand further investigation, to understand their connection with the general topic of work team assembly (size, composition, and formation), and to the (urgent) issue of software sustainability\cite{penzenstadler2013towards}. OSS is a prominent example of the ``tragedy of the commons'': companies and services benefit from the software, but there is a grossly disproportionate imbalance between those consuming the software and those building and maintaining it. Indeed, by being more aware of the internal self-organisation of their projects, owners and administrators may design strategies to optimise the collaborative efforts of the limited number (and availability) of project contributors. For instance, they can place efforts to drive the actual project's block decomposition towards a pre-defined software architectural pattern; or ensure that, despite the nested organisation within blocks, all files in a block receive some minimal attention. More research on the derivation of effective project management leadership strategies from the current division of labour in a project is clearly needed and impactful.

Closer to the complex networks and data analysis tradition, our results leave room to widen the scope of this research. To start with, future research should tackle a larger and more heterogeneous set of projects, and even across different platforms such as Bitbucket. Admittedly, this work has focused on successful projects, inasmuch we only consider a few dozens among the most popular. Beyond the richness of the analysed dataset, the relationship between maturity and structural arrangement (specially in regard to the internal organisation of subgroups) clearly demands further work. Two obvious --and intimately related-- lines of research are related to time-resolved datasets, and the design of a suitable model that can mimic the growth and evolution of OSS projects. Such model should lay down the necessary dynamical rules for both contributors and files which, presumably, differ largely. 

\section*{Material and Methods}
\paragraph{Data.} Our open source projects dataset was collected from GitHub\cite{github}, a social coding platform which provides source code management and collaboration features such as bug tracking, feature requests, tasks management and wiki for every project. Given that GitHub users can star a project (to show interest in its development and follow its advances), we chose to measure the popularity of a GitHub project in terms of its number of stars (i.e. the more stars the more popular the project is considered) and selected the 100 most popular projects. The construction of the dataset involved three phases, namely: (1) cloning, (2) import, and (3) enrichment.  

\subparagraph{Cloning and import.} After collecting the list of 100 most popular projects in GitHub (at the moment of collecting the data) via its API\cite{request}, we cloned them to collect 100 Git repositories. 
We analysed the cloned repositories and discarded those ones not involving the development of a software artifact (e.g. collection of links or questions), rejecting 15 projects out of the initial 100. We then imported the remaining Git repositories into a relational database using the Gitana~\cite{Cosentino2015} tool to facilitate the query and exploration of the projects for further analysis. In the Gitana database, Git repositories are represented in terms of users (i.e. contributors with a name and an email); files; commits (i.e. changes performed to the files); references (i.e. branches and tags); and file modifications. For two projects, the import process failed to complete due missing or corrupted information in the source GitHub repository. 

\subparagraph{Enrichment.}
Our analysis needs a clear identification of the author of each commit so that we can properly link contributors and files they have modified. Unfortunately, Git does not control the name and email contributors indicate when pushing commits resulting on clashing and duplicate problems in the data.  Clashing appears when two or more contributors have set the same name value (in Git the contributor name is manually configured), resulting in commits actually coming from different contributors appearing with the same commit name (e.g., often when using common names such as ``mike''). In addition, duplicity appears when a contributor has several emails, thus there are commits that come from the same person, but are linked to different emails suggesting different contributors. We found that, on average, around 60\% of the commits in each project were modified by contributors that involved a clashing/duplicity problem (and affecting a similar number of files). To address this problem, we relied on data provided by GitHub for each project (in particular, GitHub usernames, which are unique). By linking commits to unique usernames, we could disambiguate the contributors behind the commits. Thus, we enriched our repository data by querying GitHub API to discover the actual username for each commit in our repository, and relied on those instead on the information provided as part of the Git commit metadata. This method only failed for commits without a GitHub username associated (e.g. when the user that made that commit was no longer existing in GitHub). In those cases we stick to the email in Git commit as contributor identifier. We reduced considerably the clashing/duplicity problem in our dataset. The percentage of commits modified by contributors that may involve a clashing/duplicity problem was reduced to 0.004\% on average ($\sigma = 0.011$), and the percentage of files affected was reduced to 0.020\% ($\sigma = 0.042$). 

At the end of this process, we had successfully collected a total number of 83 projects, 48,015 contributors, 668,283 files and 912,766 commits. The other 18 projects (to the total of 65 reported in this work) were rejected due to other limitations. On one hand, we discarded some projects that presented very strong divergence between the number of nodes of the two sets, e.g. projects with very large number of files but very few contributors. In these cases, although $\altmathcal{N}$, $Q$ and $\altmathcal{I}$ can be quantified, the outcome is hardly interpretable. An example of this is the project \textit{material-designs-icons}, with 15 contributors involved in the development of 12,651 files. On the other hand, we considered only projects with a bipartite network size within the range $10^{1} \le S \le 10^4$, as the computational costs to optimise in-block nestedness and modularity for larger sizes were too severe.


\paragraph{Matrix generation.}  We build a bipartite unweighted network as a rectangular $N \times M$ matrix, where rows and columns refer to contributors and source files of an OSS project, respectively. Cells therefore represent links in the bipartite network, i.e. if the cell $a_{ij}$ has a value of $1$, it represents that the contributor $i$ has modified the file $j$ at least once, otherwise $a_{ij}$ is set to $0$.

\paragraph{Nestedness.} 
In structural terms, a nested pattern is observed when specialists (nodes with low connectivity) interact with proper nested subsets of those species interacting with generalists (nodes with high connectivity), see Figure \ref{fig_structures_matrices} (left). Several works have shown that a nested configuration is  signature feature of cooperative environments --those in which interacting species obtain some benefit\cite{bascompte2003nested,bastolla2009architecture,suweis2013emergence}. Following this example in natural systems, scholars have sought (and found) this pattern in other kinds of systems\cite{saavedra2011strong,kamilar2014cultural,borge2017emergence}. Here, we quantify the amount of nestedness in our OSS networks by employing the global nestedness fitness $\altmathcal{N}$ introduced by Sol\'e-Ribalta {\it et al.}\cite{sole2018revealing}:
\begin{center}
\begin{equation} 
\altmathcal{{N}} = \frac{2}{N+M} \left \{  \sum_{i,j}^{N} \left[  \frac{\mathit{O}_{i,j}- \langle \mathit{O}_{i,j}\rangle}{k_{j}(N - 1)} \Theta( k_{i} -k_{j} )\right]  + \sum_{l,m}^{M} \left[  \frac{\mathit{O}_{l,m}- \langle \mathit{O}_{l,m}\rangle}{k_{m}(M - 1)} \Theta( k_{l} -k_{m} )\right]  \right \},
 \label{eq_nest}
 \end{equation}
  \end{center}
where $O_{i,j}$ (or  $O_{l,m}$) measures the degree of links overlap between rows (or columns) node pairs; $k_i$,$k_j$ corresponds to the degree of the nodes $i$,$j$; $\Theta(\cdot)$ is a Heaviside step function that guarantees that we only compute the overlap between pair of nodes when $k_i \geq  k_j $. Finally, $\langle \mathit{O}_{i,j}\rangle$ represents the expected number of links between row nodes $i$ and $j$ in the null model, and is equal to $\langle \mathit{O}_{i,j}\rangle=\frac{k_ik_j}{M}$. 

\paragraph{Modularity.} A modular network structure (Figure \ref{fig_structures_matrices}, center) implies the existence of well-connected subgroups, which can be identified given the right heuristics to do so. Modularity has been reported in almost any kind of systems: from food-webs \cite{stouffer2011compartmentalization} to lexical networks \cite{borge2009navigating}, to the Internet \cite{fortunato2010community} and social networks. 

A correct quantification is needed to settle the extent to which a given network does or does not have community structure. First, we look for the optimal modular partition of the nodes through a community detection analysis\cite{fortunato2010community,barber2007modularity}. To this end, we apply the extremal optimisation algorithm\cite{duch2005community} (along with a Kernighan-Lin\cite{Kernighan1970} refinement procedure) to maximise Barber's\cite{barber2007modularity} modularity $Q$,

%
%
%
%

\begin{center}
\begin{equation} 
Q = \frac{1}{L} \displaystyle\sum^{N}_{i=1}\sum^{N+M}_{j=N+1} \left(\tilde{a}_{ij}-\tilde{p}_{ij}\right)\delta(\alpha_i,\alpha_j)
  \end{equation}
  \end{center}
where $L$ is the number of interactions (links) in the network, $\tilde{a}_{ij}$ denotes the existence of a link between nodes $i$ and $j$, $\tilde{p}_{ij}=k_{i}k_{j}/L$ is the probability that a link exists by chance, and $\delta(\alpha_i,\alpha_j)$ is the Kronecker delta function, which takes the value 1 if nodes $i$ and $j$ are in the same community, and 0 otherwise. 

\paragraph{In-block nestedness.} The third architectural pattern that we considered in our work, consists of a mesoscale hybrid pattern in which the network presents a modular structure, but the interactions within each module are nested, i.e., an in-block nested structure Figure \ref{fig_structures_matrices} (right). This type of hybrid or ``compound" architectures, were first described by Lewinsohn {\it et al}.\cite{lewinsohn2006structure}. Although, the literature covering this types of patterns is still scarce, the existence of such type of hybrid structure in empirical networks has been recently explored \cite{flores2013multi,beckett2013coevolutionary,sole2018revealing} and the results from these works seem to indicate that combined structures are in fact, a common feature in many systems from different contexts.

In order to compute the amount of in-block nested present in networks, in this work, we have adopted a new objective function\cite{sole2018revealing}, that is capable to detect these hybrid architectures, and employed the same optimisation algorithms used to maximise modularity. The in-block nestedness objective function can be written as,

\begin{center}
\begin{equation}
\altmathcal{I} = \frac{2}{N+M} \left \{  \sum_{i,j}^{N} \left[  \frac{\mathit{O}_{i,j}- \langle \mathit{O}_{i,j}\rangle}{k_{j}(C_{i} - 1)} \Theta( k_{i} -k_{j} ) \delta (\alpha _{i}, \alpha _{j}) \right] +  \sum_{l,m}^{M} \left[  \frac{\mathit{O}_{l,m}- \langle \mathit{O}_{l,m}\rangle}{k_{m}(C_{l} - 1)} \Theta( k_{l} -k_{m} ) \delta (\alpha _{l}, \alpha _{m}) \right] \right \},
\label{eq_inb}
\end{equation}
\end{center}

Note that, by definition, $\altmathcal{I}$ reduces to $\altmathcal{N}$ when the number of blocks is 1. This explains why the right half of the ternary plot (Figure~\ref{fig_ternary}) is necessarily empty: $\altmathcal{I} \ge \altmathcal{N}$, and therefore $f_{\altmathcal{I}} \ge f_{\altmathcal{N}}$. On the other hand, an in-block nested structure exhibits necessarily some level of modularity, but not the other way around. This explains why the lower-left area of the simplex in Figure~\ref{fig_ternary} is empty as well (see Palazzi {\it et al.}\cite{palazzi2018antagonistic} for details).


\begin{thebibliography}{10}
\expandafter\ifx\csname url\endcsname\relax
  \def\url#1{\texttt{#1}}\fi
\expandafter\ifx\csname urlprefix\endcsname\relax\def\urlprefix{URL }\fi
\expandafter\ifx\csname doiprefix\endcsname\relax\def\doiprefix{DOI }\fi
\providecommand{\bibinfo}[2]{#2}
\providecommand{\eprint}[2][]{\url{#2}}

\bibitem{opensource}
\bibinfo{title}{Open source initiative}.
\newblock \bibinfo{howpublished}{\url{https://opensource.org/}}.

\bibitem{SchuwerGH15}
\bibinfo{author}{Schuwer, R.}, \bibinfo{author}{van Genuchten, M.} \&
  \bibinfo{author}{Hatton, L.}
\newblock \bibinfo{journal}{\bibinfo{title}{On the impact of being open}}.
\newblock {\emph{\JournalTitle{{IEEE} Software}}}
  \textbf{\bibinfo{volume}{32}}, \bibinfo{pages}{81--83}
  (\bibinfo{year}{2015}).

\bibitem{Dabbish2012}
\bibinfo{author}{Dabbish, L.}, \bibinfo{author}{Stuart, C.},
  \bibinfo{author}{Tsay, J.} \& \bibinfo{author}{Herbsleb, J.}
\newblock \bibinfo{title}{{Social Coding in {G}it{H}ub: Transparency and
  Collaboration in an Open Software Repository}}.
\newblock In \emph{\bibinfo{booktitle}{ACM Conf. on Computer-Supported
  Cooperative Work and Social Computing}}, \bibinfo{pages}{1277--1286}
  (\bibinfo{year}{2012}).

\bibitem{Padhye2014}
\bibinfo{author}{Padhye, R.}, \bibinfo{author}{Mani, S.} \&
  \bibinfo{author}{Sinha, V.~S.}
\newblock \bibinfo{title}{{A Study of External Community Contribution to
  Open-source Projects on {G}it{H}ub}}.
\newblock In \emph{\bibinfo{booktitle}{Working Conf. on Mining Software
  Repositories}}, \bibinfo{pages}{332--335} (\bibinfo{year}{2014}).

\bibitem{Lima2014}
\bibinfo{author}{Lima, A.}, \bibinfo{author}{Rossi, L.} \&
  \bibinfo{author}{Musolesi, M.}
\newblock \bibinfo{title}{{Coding Together at Scale: {G}it{H}ub as a
  Collaborative Social Network}}.
\newblock In \emph{\bibinfo{booktitle}{Int. Conf. on Weblogs and Social
  Media}}, \bibinfo{pages}{10} (\bibinfo{year}{2014}).

\bibitem{Dabbish2013}
\bibinfo{author}{Dabbish, L.}, \bibinfo{author}{Stuart, C.},
  \bibinfo{author}{Tsay, J.} \& \bibinfo{author}{Herbsleb, J.}
\newblock \bibinfo{journal}{\bibinfo{title}{{Leveraging Transparency}}}.
\newblock {\emph{\JournalTitle{{IEEE} Software}}}
  \textbf{\bibinfo{volume}{30}}, \bibinfo{pages}{37--43}
  (\bibinfo{year}{2013}).

\bibitem{Fitz-Gerald12b}
\bibinfo{author}{Fitz{-}Gerald, S.}
\newblock \bibinfo{journal}{\bibinfo{title}{Schweik and english, 2012 -
  internet success: {A} study of open-source software commons, {C.M.} schweik,
  {R.C.} english. {MIT} press {(2012)}}}.
\newblock {\emph{\JournalTitle{Int. Journal on Information Management}}}
  \textbf{\bibinfo{volume}{32}}, \bibinfo{pages}{596--597}
  (\bibinfo{year}{2012}).

\bibitem{CosentinoIC15}
\bibinfo{author}{Cosentino, V.}, \bibinfo{author}{Izquierdo, J. L.~C.} \&
  \bibinfo{author}{Cabot, J.}
\newblock \bibinfo{title}{Assessing the bus factor of git repositories}.
\newblock In \emph{\bibinfo{booktitle}{Int. Conf. on Software Analysis,
  Evolution, and Reengineering}}, \bibinfo{pages}{499--503}
  (\bibinfo{year}{2015}).

\bibitem{Yamashita2015}
\bibinfo{author}{Yamashita, K.}, \bibinfo{author}{McIntosh, S.},
  \bibinfo{author}{Kamei, Y.}, \bibinfo{author}{Hassan, A.~E.} \&
  \bibinfo{author}{Ubayashi, N.}
\newblock \bibinfo{title}{{Revisiting the Applicability of the Pareto Principle
  to Core Development Teams in Open Source Software Projects}}.
\newblock In \emph{\bibinfo{booktitle}{Int. Workshop on Principles of Software
  Evolution}}, \bibinfo{pages}{46--55} (\bibinfo{year}{2015}).

\bibitem{Avelino2016}
\bibinfo{author}{Avelino, G.}, \bibinfo{author}{Passos, L.},
  \bibinfo{author}{Hora, A.} \& \bibinfo{author}{Valente, M.~T.}
\newblock \bibinfo{title}{A novel approach for estimating truck factors}.
\newblock In \emph{\bibinfo{booktitle}{Int. Conf. on Program Comprehension}},
  \bibinfo{pages}{1--10} (\bibinfo{year}{2016}).

\bibitem{Pham2013}
\bibinfo{author}{Pham, R.}, \bibinfo{author}{Singer, L.},
  \bibinfo{author}{Liskin, O.}, \bibinfo{author}{Figueira~Filho, F.} \&
  \bibinfo{author}{Schneider, K.}
\newblock \bibinfo{title}{{Creating a Shared Understanding of Testing Culture
  on a Social Coding Site}}.
\newblock In \emph{\bibinfo{booktitle}{Int. Conf. on Software Engineering}},
  \bibinfo{pages}{112--121} (\bibinfo{year}{2013}).

\bibitem{Yamashita2016}
\bibinfo{author}{Yamashita, K.}, \bibinfo{author}{Kamei, Y.},
  \bibinfo{author}{McIntosh, S.}, \bibinfo{author}{Hassan, A.~E.} \&
  \bibinfo{author}{Ubayashi, N.}
\newblock \bibinfo{journal}{\bibinfo{title}{{Magnet or Sticky? Measuring
  Project Characteristics from the Perspective of Developer Attraction and
  Retention}}}.
\newblock {\emph{\JournalTitle{Journal of Information Processing}}}
  \textbf{\bibinfo{volume}{24}}, \bibinfo{pages}{339--348}
  (\bibinfo{year}{2016}).

\bibitem{Hata2015}
\bibinfo{author}{Hata, H.}, \bibinfo{author}{Todo, T.}, \bibinfo{author}{Onoue,
  S.} \& \bibinfo{author}{Matsumoto, K.}
\newblock \bibinfo{title}{{Characteristics of Sustainable OSS Projects: a
  Theoretical and Empirical Study}}.
\newblock In \emph{\bibinfo{booktitle}{Int. Workshop on Cooperative and Human
  Aspects of Software Engineering}}, \bibinfo{pages}{15--21}
  (\bibinfo{year}{2015}).

\bibitem{BertholdoG16}
\bibinfo{author}{Bertholdo, A. P.~O.} \& \bibinfo{author}{Gerosa, M.~A.}
\newblock \bibinfo{title}{{Promoting Engagement in Open Collaboration
  Communities by Means of Gamification}}.
\newblock In \emph{\bibinfo{booktitle}{Int. Conf. on Human-Computer
  Interaction}}, \bibinfo{pages}{15--20} (\bibinfo{year}{2016}).

\bibitem{SteinmacherCTG16}
\bibinfo{author}{Steinmacher, I.}, \bibinfo{author}{Conte, T.~U.},
  \bibinfo{author}{Treude, C.} \& \bibinfo{author}{Gerosa, M.~A.}
\newblock \bibinfo{title}{Overcoming open source project entry barriers with a
  portal for newcomers}.
\newblock In \emph{\bibinfo{booktitle}{Int. Conf. on Software Engineering}},
  \bibinfo{pages}{273--284} (\bibinfo{year}{2016}).

\bibitem{SteinmacherSGR15}
\bibinfo{author}{Steinmacher, I.}, \bibinfo{author}{Silva, M. A.~G.},
  \bibinfo{author}{Gerosa, M.~A.} \& \bibinfo{author}{Redmiles, D.~F.}
\newblock \bibinfo{journal}{\bibinfo{title}{A systematic literature review on
  the barriers faced by newcomers to open source software projects}}.
\newblock {\emph{\JournalTitle{Information {\&} Software Technology}}}
  \textbf{\bibinfo{volume}{59}}, \bibinfo{pages}{67--85}
  (\bibinfo{year}{2015}).

\bibitem{cosentino2017systematic}
\bibinfo{author}{Cosentino, V.}, \bibinfo{author}{Izquierdo, J. L.~C.} \&
  \bibinfo{author}{Cabot, J.}
\newblock \bibinfo{journal}{\bibinfo{title}{A systematic mapping study of
  software development with github}}.
\newblock {\emph{\JournalTitle{IEEE Access}}} \textbf{\bibinfo{volume}{5}},
  \bibinfo{pages}{7173--7192} (\bibinfo{year}{2017}).

\bibitem{valverde2007self}
\bibinfo{author}{Valverde, S.} \& \bibinfo{author}{Sol{\'e}, R.~V.}
\newblock \bibinfo{journal}{\bibinfo{title}{Self-organization versus hierarchy
  in open-source social networks}}.
\newblock {\emph{\JournalTitle{Physical Review E}}}
  \textbf{\bibinfo{volume}{76}}, \bibinfo{pages}{046118}
  (\bibinfo{year}{2007}).

\bibitem{bird2008latent}
\bibinfo{author}{Bird, C.}, \bibinfo{author}{Pattison, D.},
  \bibinfo{author}{D'Souza, R.}, \bibinfo{author}{Filkov, V.} \&
  \bibinfo{author}{Devanbu, P.}
\newblock \bibinfo{title}{Latent social structure in open source projects}.
\newblock In \emph{\bibinfo{booktitle}{Proceedings of the 16th ACM SIGSOFT
  International Symposium on Foundations of software engineering}},
  \bibinfo{pages}{24--35} (\bibinfo{year}{2008}).

\bibitem{hong2011understanding}
\bibinfo{author}{Hong, Q.}, \bibinfo{author}{Kim, S.}, \bibinfo{author}{Cheung,
  S.~C.} \& \bibinfo{author}{Bird, C.}
\newblock \bibinfo{title}{Understanding a developer social network and its
  evolution}.
\newblock In \emph{\bibinfo{booktitle}{Software Maintenance (ICSM), 2011 27th
  IEEE International Conference on}}, \bibinfo{pages}{323--332}
  (\bibinfo{organization}{IEEE}, \bibinfo{year}{2011}).

\bibitem{dunbar1992neocortex}
\bibinfo{author}{Dunbar, R.}
\newblock \bibinfo{journal}{\bibinfo{title}{Neocortex size as a constraint on
  group size in primates}}.
\newblock {\emph{\JournalTitle{Journal of Human Evolution}}}
  \textbf{\bibinfo{volume}{22}}, \bibinfo{pages}{469--493}
  (\bibinfo{year}{1992}).

\bibitem{gonccalves2011modeling}
\bibinfo{author}{Gon{\c{c}}alves, B.}, \bibinfo{author}{Perra, N.} \&
  \bibinfo{author}{Vespignani, A.}
\newblock \bibinfo{journal}{\bibinfo{title}{Modeling users' activity on twitter
  networks: Validation of dunbar's number}}.
\newblock {\emph{\JournalTitle{PloS One}}} \textbf{\bibinfo{volume}{6}},
  \bibinfo{pages}{e22656} (\bibinfo{year}{2011}).

\bibitem{patterson1986patterson}
\bibinfo{author}{Patterson, B.~D.} \& \bibinfo{author}{Atmar, W.}
\newblock \bibinfo{journal}{\bibinfo{title}{Nested subsets and the structure of
  insular mammalian faunas and archipelagos}}.
\newblock {\emph{\JournalTitle{Biological Journal of the Linnean Society}}}
  \textbf{\bibinfo{volume}{28}}, \bibinfo{pages}{65--82}
  (\bibinfo{year}{1986}).

\bibitem{atmar1993measureorder}
\bibinfo{author}{Atmar, W.} \& \bibinfo{author}{Patterson, B.~D.}
\newblock \bibinfo{journal}{\bibinfo{title}{The measure of order and disorder
  in the distribution of species in fragmented habitat}}.
\newblock {\emph{\JournalTitle{Oecologia}}} \textbf{\bibinfo{volume}{96}},
  \bibinfo{pages}{373--382} (\bibinfo{year}{1993}).

\bibitem{newman2004finding}
\bibinfo{author}{Newman, M.~E.} \& \bibinfo{author}{Girvan, M.}
\newblock \bibinfo{journal}{\bibinfo{title}{Finding and evaluating community
  structure in networks}}.
\newblock {\emph{\JournalTitle{Physical Review E}}}
  \textbf{\bibinfo{volume}{69}}, \bibinfo{pages}{026113}
  (\bibinfo{year}{2004}).

\bibitem{barber2007modularity}
\bibinfo{author}{Barber, M.~J.}
\newblock \bibinfo{journal}{\bibinfo{title}{Modularity and community detection
  in bipartite networks}}.
\newblock {\emph{\JournalTitle{Physical Review E}}}
  \textbf{\bibinfo{volume}{76}}, \bibinfo{pages}{066102}
  (\bibinfo{year}{2007}).

\bibitem{fortunato2010community}
\bibinfo{author}{Fortunato, S.}
\newblock \bibinfo{journal}{\bibinfo{title}{Community detection in graphs}}.
\newblock {\emph{\JournalTitle{Physics Reports}}}
  \textbf{\bibinfo{volume}{486}}, \bibinfo{pages}{75--174}
  (\bibinfo{year}{2010}).

\bibitem{Brooks1995}
\bibinfo{author}{Brooks, F.~P.}
\newblock \emph{\bibinfo{title}{{The Mythical Man-Month - Essays on Software
  Engineering {(2.} ed.)}}} (\bibinfo{publisher}{Addison-Wesley},
  \bibinfo{year}{1995}).

\bibitem{lewinsohn2006structure}
\bibinfo{author}{Lewinsohn, T.~M.}, \bibinfo{author}{In{\'a}cio~Prado, P.},
  \bibinfo{author}{Jordano, P.}, \bibinfo{author}{Bascompte, J.} \&
  \bibinfo{author}{Olesen, J.~M.}
\newblock \bibinfo{journal}{\bibinfo{title}{Structure in plant--animal
  interaction assemblages}}.
\newblock {\emph{\JournalTitle{Oikos}}} \textbf{\bibinfo{volume}{113}},
  \bibinfo{pages}{174--184} (\bibinfo{year}{2006}).

\bibitem{flores2013multi}
\bibinfo{author}{Flores, C.~O.}, \bibinfo{author}{Valverde, S.} \&
  \bibinfo{author}{Weitz, J.~S.}
\newblock \bibinfo{journal}{\bibinfo{title}{Multi-scale structure and
  geographic drivers of cross-infection within marine bacteria and phages}}.
\newblock {\emph{\JournalTitle{The ISME journal}}}
  \textbf{\bibinfo{volume}{7}}, \bibinfo{pages}{520--532}
  (\bibinfo{year}{2013}).

\bibitem{sole2018revealing}
\bibinfo{author}{Sol{\'e}-Ribalta, A.}, \bibinfo{author}{Tessone, C.~J.},
  \bibinfo{author}{Mariani, M.~S.} \& \bibinfo{author}{Borge-Holthoefer, J.}
\newblock \bibinfo{journal}{\bibinfo{title}{Revealing in-block nestedness:
  detection and benchmarking}}.
\newblock {\emph{\JournalTitle{Physical Review E}}}
  \textbf{\bibinfo{volume}{97}}, \bibinfo{pages}{062302}
  (\bibinfo{year}{2018}).

\bibitem{palazzi2018antagonistic}
\bibinfo{author}{Palazzi, M.}, \bibinfo{author}{Borge-Holthoefer, J.},
  \bibinfo{author}{Tessone, C.} \& \bibinfo{author}{Sol{\'e}-Ribalta, A.}
\newblock \bibinfo{journal}{\bibinfo{title}{Antagonistic structural patterns in
  complex networks}}.
\newblock {\emph{\JournalTitle{arXiv preprint arXiv:1810.12785}}}
  (\bibinfo{year}{2018}).

\bibitem{dunbar2010many}
\bibinfo{author}{Dunbar, R.}
\newblock \emph{\bibinfo{title}{How many friends does one person need?:
  Dunbar's number and other evolutionary quirks}} (\bibinfo{publisher}{Faber \&
  Faber}, \bibinfo{year}{2010}).

\bibitem{derex2016partial}
\bibinfo{author}{Derex, M.} \& \bibinfo{author}{Boyd, R.}
\newblock \bibinfo{journal}{\bibinfo{title}{Partial connectivity increases
  cultural accumulation within groups}}.
\newblock {\emph{\JournalTitle{Proceedings of the National Academy of
  Sciences}}} \textbf{\bibinfo{volume}{113}}, \bibinfo{pages}{2982--2987}
  (\bibinfo{year}{2016}).

\bibitem{derex2018divide}
\bibinfo{author}{Derex, M.}, \bibinfo{author}{Perreault, C.} \&
  \bibinfo{author}{Boyd, R.}
\newblock \bibinfo{journal}{\bibinfo{title}{Divide and conquer: intermediate
  levels of population fragmentation maximize cultural accumulation}}.
\newblock {\emph{\JournalTitle{Phil. Trans. R. Soc. B}}}
  \textbf{\bibinfo{volume}{373}}, \bibinfo{pages}{20170062}
  (\bibinfo{year}{2018}).

\bibitem{borge2017emergence}
\bibinfo{author}{Borge-Holthoefer, J.}, \bibinfo{author}{Ba{\~n}os, R.~A.},
  \bibinfo{author}{Gracia-L{\'a}zaro, C.} \& \bibinfo{author}{Moreno, Y.}
\newblock \bibinfo{journal}{\bibinfo{title}{Emergence of consensus as a
  modular-to-nested transition in communication dynamics}}.
\newblock {\emph{\JournalTitle{Scientific Reports}}}
  \textbf{\bibinfo{volume}{7}}, \bibinfo{pages}{41673} (\bibinfo{year}{2017}).

\bibitem{penzenstadler2013towards}
\bibinfo{author}{Penzenstadler, B.}
\newblock \bibinfo{title}{Towards a definition of sustainability in and for
  software engineering}.
\newblock In \emph{\bibinfo{booktitle}{Proceedings of the 28th Annual ACM
  Symposium on Applied Computing}}, \bibinfo{pages}{1183--1185}
  (\bibinfo{year}{2013}).

\bibitem{github}
\bibinfo{title}{\url{https://github.com}}.

\bibitem{request}
\bibinfo{title}{Using the request:
  \url{https://api.github.com/search/repositories?q=stars:>1\&sort=stars\&order=desc\&per\_page=100}}.

\bibitem{Cosentino2015}
\bibinfo{author}{Cosentino, V.}, \bibinfo{author}{{C{\'{a}}novas Izquierdo},
  J.~L.} \& \bibinfo{author}{Cabot, J.}
\newblock \bibinfo{title}{{Gitana: {A} SQL-Based Git Repository Inspector}}.
\newblock In \emph{\bibinfo{booktitle}{Int. Conf. on Conceptual Modeling}},
  \bibinfo{pages}{329--343} (\bibinfo{year}{2015}).

\bibitem{bascompte2003nested}
\bibinfo{author}{Bascompte, J.}, \bibinfo{author}{Jordano, P.},
  \bibinfo{author}{Meli{\'a}n, C.~J.} \& \bibinfo{author}{Olesen, J.~M.}
\newblock \bibinfo{journal}{\bibinfo{title}{The nested assembly of
  plant--animal mutualistic networks}}.
\newblock {\emph{\JournalTitle{Proceedings of the National Academy of
  Sciences}}} \textbf{\bibinfo{volume}{100}}, \bibinfo{pages}{9383--9387}
  (\bibinfo{year}{2003}).

\bibitem{bastolla2009architecture}
\bibinfo{author}{Bastolla, U.} \emph{et~al.}
\newblock \bibinfo{journal}{\bibinfo{title}{The architecture of mutualistic
  networks minimizes competition and increases biodiversity}}.
\newblock {\emph{\JournalTitle{Nature}}} \textbf{\bibinfo{volume}{458}},
  \bibinfo{pages}{1018--1020} (\bibinfo{year}{2009}).

\bibitem{suweis2013emergence}
\bibinfo{author}{Suweis, S.}, \bibinfo{author}{Simini, F.},
  \bibinfo{author}{Banavar, J.~R.} \& \bibinfo{author}{Maritan, A.}
\newblock \bibinfo{journal}{\bibinfo{title}{Emergence of structural and
  dynamical properties of ecological mutualistic networks}}.
\newblock {\emph{\JournalTitle{Nature}}} \textbf{\bibinfo{volume}{500}},
  \bibinfo{pages}{449} (\bibinfo{year}{2013}).

\bibitem{saavedra2011strong}
\bibinfo{author}{Saavedra, S.}, \bibinfo{author}{Stouffer, D.~B.},
  \bibinfo{author}{Uzzi, B.} \& \bibinfo{author}{Bascompte, J.}
\newblock \bibinfo{journal}{\bibinfo{title}{{Strong Contributors to Network
  Persistence Are the Most Vulnerable to Extinction}}}.
\newblock {\emph{\JournalTitle{Nature}}} \textbf{\bibinfo{volume}{478}},
  \bibinfo{pages}{233--235} (\bibinfo{year}{2011}).

\bibitem{kamilar2014cultural}
\bibinfo{author}{Kamilar, J.~M.} \& \bibinfo{author}{Atkinson, Q.~D.}
\newblock \bibinfo{journal}{\bibinfo{title}{Cultural assemblages show nested
  structure in humans and chimpanzees but not orangutans}}.
\newblock {\emph{\JournalTitle{Proceedings of the National Academy of
  Sciences}}} \textbf{\bibinfo{volume}{111}}, \bibinfo{pages}{111--115}
  (\bibinfo{year}{2014}).

\bibitem{stouffer2011compartmentalization}
\bibinfo{author}{Stouffer, D.~B.} \& \bibinfo{author}{Bascompte, J.}
\newblock \bibinfo{journal}{\bibinfo{title}{Compartmentalization increases
  food-web persistence}}.
\newblock {\emph{\JournalTitle{Proceedings of the National Academy of
  Sciences}}} \textbf{\bibinfo{volume}{108}}, \bibinfo{pages}{3648--3652}
  (\bibinfo{year}{2011}).

\bibitem{borge2009navigating}
\bibinfo{author}{Borge-Holthoefer, J.} \& \bibinfo{author}{Arenas, A.}
\newblock \bibinfo{title}{{Navigating Word Association Norms to Extract
  Semantic Information}}.
\newblock In \emph{\bibinfo{booktitle}{An. Conf. of the Cognitive Science
  Society}} (\bibinfo{year}{2009}).

\bibitem{duch2005community}
\bibinfo{author}{Duch, J.} \& \bibinfo{author}{Arenas, A.}
\newblock \bibinfo{journal}{\bibinfo{title}{Community detection in complex
  networks using extremal optimization}}.
\newblock {\emph{\JournalTitle{Physical Review E}}}
  \textbf{\bibinfo{volume}{72}}, \bibinfo{pages}{027104}
  (\bibinfo{year}{2005}).

\bibitem{Kernighan1970}
\bibinfo{author}{Kernighan, B.~W.} \& \bibinfo{author}{Lin, S.}
\newblock \bibinfo{journal}{\bibinfo{title}{An efficient heuristic procedure
  for partitioning graphs}}.
\newblock {\emph{\JournalTitle{The Bell system technical journal}}}
  \textbf{\bibinfo{volume}{49}}, \bibinfo{pages}{291--307}
  (\bibinfo{year}{1970}).

\bibitem{beckett2013coevolutionary}
\bibinfo{author}{Beckett, S.~J.} \& \bibinfo{author}{Williams, H.~T.}
\newblock \bibinfo{journal}{\bibinfo{title}{Coevolutionary diversification
  creates nested-modular structure in phage--bacteria interaction networks}}.
\newblock {\emph{\JournalTitle{Interface Focus}}} \textbf{\bibinfo{volume}{3}},
  \bibinfo{pages}{20130033} (\bibinfo{year}{2013}).

\end{thebibliography}


\section*{Author contributions statement}

All authors designed research. M.J.P., A.S-R. and J.B-H. and performed research. All authors analysed the results. J.B-H. and A.S-R. wrote the paper. All authors approved the final version.

\noindent\textbf{Competing Interests} The authors declare no competing interests.

\noindent\textbf{Correspondence} Correspondence and requests for materials should be addressed to J.B-H. \\ (email: jborgeh@uoc.edu).

\end{document}